\documentclass[runningheads]{llncs}
\usepackage{bbding}
\usepackage{graphicx}
\usepackage[marginal]{footmisc}

\begin{document}
\title{Hepatocellular Carcinoma Segmentation from Digital Subtraction Angiography Videos using Learnable Temporal Difference\thanks{The work was supported in part by Key-Area Research and Development Program of Guangdong Province [2020B0101350001], in part by the National Key R\&D Program of China with grant No. 2018YFB1800800, by Shenzhen Outstanding Talents Training Fund, and by Guangdong Research Project No. 2017ZT07X152. It was also supported by NFSC 61931024 and National Natural Science Foundation of China (81871323).}}

\titlerunning{HCC Segmentation in DSA videos}


\author{Wenting Jiang\inst{1,2} \and 
Yicheng Jiang\inst{1,2} \and 
Lu Zhang\inst{3} \and 
Changmiao Wang\inst{2} \and 
Xiaoguang Han\inst{1,2(}\Envelope\inst{)} \and 
Shuixing Zhang\inst{3(}\Envelope\inst{)} \and 
Xiang Wan\inst{2} \and 
Shuguang Cui\inst{1,2,4} 
}

\authorrunning{W. Jiang et al.}
%


\institute{
School of Science and Engineering, \\ The Chinese University of Hong Kong, Shenzhen, Shenzhen, China \and
Shenzhen Research Institute of Big Data, Shenzhen, China \\
\email{hanxiaoguang@cuhk.edu.cn} \and
Department of Radiology, The First Affiliated Hospital of Jinan University, Guangzhou, China \\
\email{shui7515@126.com} \and
Future Network of Intelligence Institue, Shenzhen, China
}

\maketitle

\footnote{Wenting Jiang, Yicheng Jiang and Lu Zhang contribute equally to this work.\\
The corresponding authors are Xiaoguang Han and Shuixing Zhang.
}

\begin{abstract}
Automatic segmentation of hepatocellular carcinoma (HCC) in Digital Subtraction Angiography (DSA) videos can assist radiologists in efficient diagnosis of HCC and accurate evaluation of tumors in clinical practice. Few studies have investigated HCC segmentation from DSA videos. It shows great challenging due to motion artifacts in filming, ambiguous boundaries of tumor regions and high similarity in imaging to other anatomical tissues. In this paper, we raise the problem of HCC segmentation in DSA videos, and build our own DSA dataset. We also propose a novel segmentation network called DSA-LTDNet, including a segmentation sub-network, a temporal difference learning (TDL) module and a liver region segmentation (LRS) sub-network for providing additional guidance. DSA-LTDNet is preferable for learning the latent motion information from DSA videos proactively and boosting segmentation performance. All of experiments are conducted on our self-collected dataset. Experimental results show that DSA-LTDNet increases the DICE score by nearly 4\% compared to the U-Net baseline.

\end{abstract}

\section{Introduction}

\begin{figure}[t]
  \centering
  \includegraphics[width=1.0\columnwidth]{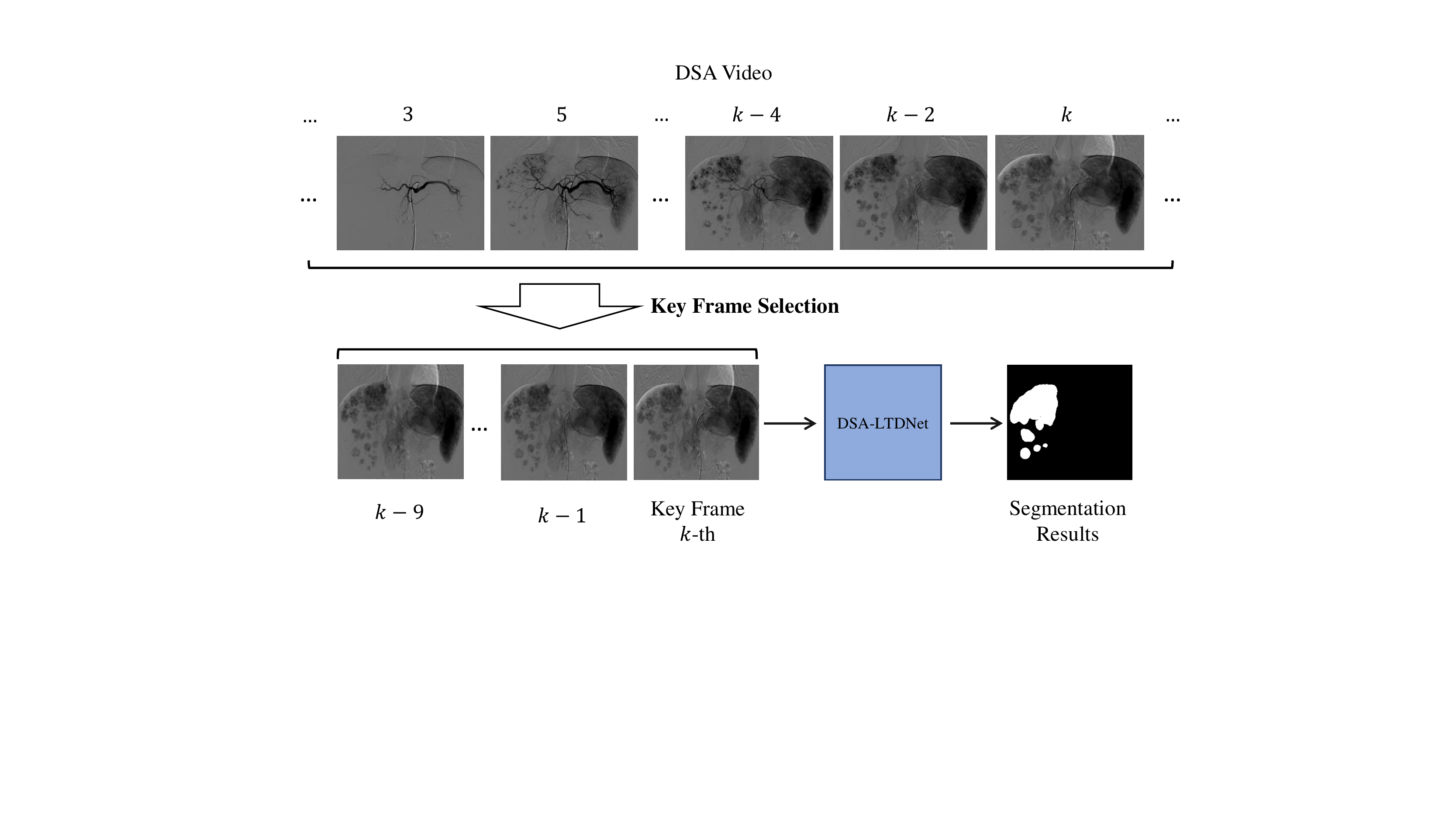}
  \caption{Samples of DSA video frames and the pipeline of our methodology.}
  \label{fig1}
\end{figure}

Digital Subtraction Angiography (DSA) is good at displaying vascular lesions, thus it is widely used in clinical detection of blood flow, stenosis, thrombosis, etc. DSA has been one of the mainstream medical imaging tools for evaluating blood supply of tumors in Transcatheter Arterial Chemoembolization (TACE) for Hepatocellular Carcinoma (HCC) \cite{liapi2007transcatheter}. It is also essential for quantitative analysis of tumor blood supply in clinical diagnosis, which provides important information and guidance for operative drug planning and postoperative prognostic. In particular, radiologists are usually required to accurately identify multiple tumors of various sizes, healthy anatomical tissues and motion artifacts before evaluating tumors when the imaging of tumors becomes stable in videos. Automatically segmenting HCC from DSA videos can assist radiologists in locating and delineating tumor regions quickly, which is of great value to clinical diagnosis. 

It is regretful that few studies have investigated segmentation of DSA videos before. Segmentation of DSA Videos is similar to the video object segmentation (VOS) task. There are several methods of the VOS task using traditional object detection solutions, such as optical flow, frame difference and background subtraction \cite{cheng2017segflow,guo2017new,tsai2016video}. These solutions focus on learning objects' displacement to improve object detection performance. Apparent temporal motion information of tumors also exists in DSA videos. However, different from the VOS task, the HCC segmentation aims to segment tumors on one key frame (Fig.\ref{fig1}) without inputting segmentation masks. The temporal motion information of tumors in DSA video reflects in regional expansion, color tarnish and edge sharpness over time, which is distinct from the information in the VOS task of natural scenes as well. Thus, the methods of VOS cannot be used directly in our task.

Although the HCC segmentation can be solved with classic methods of medical image segmentation, like U-Net\cite{ronneberger2015u}, the results based on these methods are not satisfactory (section 4.3). Two key issues in our problem remain unsolved: (1) Inevitable and irregular motion artifacts appearing throughout DSA videos especially those near the liver regions, slash the contrast of tumors and blur the boundary of tumors, which impairs the segmentation performance greatly. (2) Other anatomical structures (like gastrointestinal tissue filled with contrast agent) have similar shape, contrast and intensity to tumors in DSA videos leading to segmentation inaccuracy. Therefore, precise and robust segmentation is highly desired and difficult for DSA videos at the same time.

In this paper, we propose a novel segmentation network called DSA-LTDNet to address the aforementioned issues by learning the distinct motion information of tumors from DSA videos and integrating the prior anatomical knowledge of liver regions as a supplemental information. All experiments are performed on our self-collected dataset. The experimental results of our method are improved by nearly 4\% of DICE coefficient compared to the baseline. Our main contributions are: \\
1) We raise the problem of HCC segmentation in DSA videos for the first time, which is significant and challenging in clinical practice. We also build our self-collected DSA dataset of 488 samples, with careful annotations; \\
2) The creative proposed temporal difference learning network can efficiently capture the distinct motion information of tumors from DSA videos, which is helpful for delineating tumor regions;\\
3) Prior anatomical knowledge of liver regions integrated in our network can support locating tumor regions.

\section{Methodology}
\begin{figure}[ht]
  \centering
  \includegraphics[width=1\columnwidth]{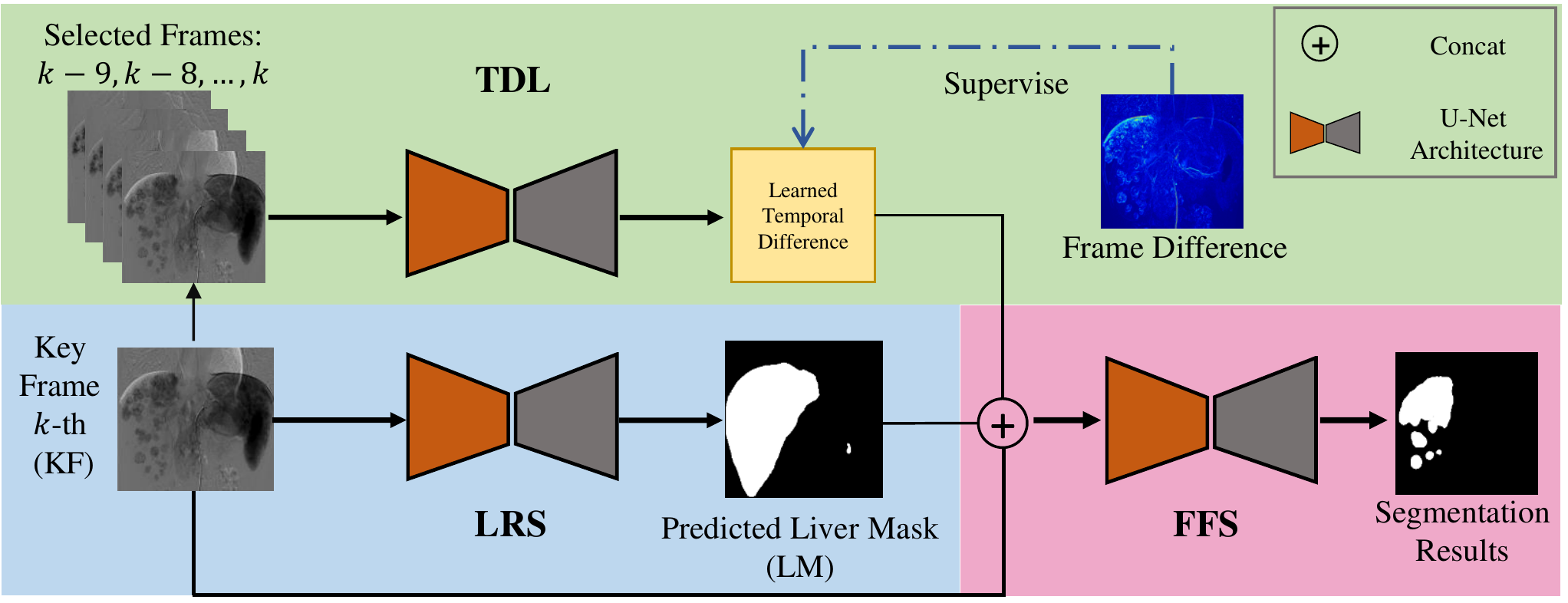}
  \caption{Architecture of DSA-LTDNet}
  \label{fig2}
\end{figure}

The pipeline of the proposed DSA-LTDNet is plotted in Fig.\ref{fig2}, including a temporal difference learning (TDL) module, a liver region segmentation (LRS) sub-network and a final fusion segmentation (FFS) sub-network. We also develop a simple approach for selecting key frames. The proposed TDL with multi-frame inputs is adopted for auto-learning temporal difference under the supervision of frame difference images. Our method also integrates prior anatomical knowledge of liver regions as guidance. The 3 inputs of FFS are the key frame, liver region masks predicted by LRS and the learned temporal difference by TDL.

\subsection{Key Frame Selection and Segmentation}
\subsubsection{Selection}
In clinical practice, radiologists choose the frame where the imaging of tumors is the clearest, the most stable and the most complete in the DSA video, for diagnosis and measurement of tumor parameters. Thus, we define such frames as the key frames and pick out them from DSA videos as a simulation of diagnosis process. We design a simple method of automatic selection for this task. Based on an observed regular pattern, we firstly calculate the difference image between 2 adjacent frames in the last 15 frames of DSA videos and sum to gain the total pixel values. To judge whether a frame is the key frame, we average the pixel values of the 2 difference images before and after the frame. We select the frame with the minimum average pixel value as the key frame because the imaging in the key frame is the most stable imaging, representing the minimum average pixel value. 

\subsubsection{Segmentation}
Several classical medical image segmentation networks using the single key frame as the input are tested. Their segmentation results on the key frames (section 4.2.) are similar to each other. Their overall performances are barely satisfactory due to the challenging work. We finally choose the basic U-Net as the baseline because it is the easiest one to implement and modify.

\subsection{Temporal Difference Learning}

\subsubsection{Motion Information} 
The result of the U-Net baseline on the key frames reflect many problems, including multiple false positive and negative regions in the segmentation mask (section 4.3). However, the motion information in the frames, such as the pixel-based frame difference (Fig.\ref{fig1}), can help delineate the tumor regions better. Consequently, we propose to add motion information across frames into our network for superior segmentation performance. We attempted 3 classic VOS methods of extracting motion information, which are the optical flow \cite{shafie2009motion}, frame difference \cite{singla2014motion} and background subtraction difference \cite{kavitha2012vibe}. Our experiments (section 4.2 and Table \ref{tab2}) show that FSS with inputs, key frames(KF) concating frame differences(FD) performs best(71.73\%). FDs are calculated between the $(k-9)-th$ and the $k-th$ for the purpose of acquiring the most representative difference from the $k-th$ key frame (Fig.\ref{fig2}). We choose the parameter, 9, based on our experimental results with different parameters: 5, 7, 9, 11, 13. The model with k-9 has the best result due to its bigger difference than that of 5, 7 and less noise than that of 11, 13 according to our analysis.

\subsubsection{Learnable Difference} 
Such motion information as FD is simple and direct, while humans can obtain more complex motion information. Thus, we also propose a Temporal Difference Learning (TDL) module, to better learn the latent temporal difference automatically. TDL module concatenates the 10 consecutive frames (Fig.\ref{fig2}) as its input and uses U-Net to learn the latent motion information from the input. We believe that this module has the ability to dig out more useful motion information than using the frame difference directly. However, the results in section 4.2 show that TDL cannot perform well without any supervision because it is difficult to converge. Therefore, we creatively use the frame difference as ancillary information to supervise TDL. Specifically, we creatively initialize TDL guided by FD, and finetune it with a relatively small loss in DSA-LTDNet. Experimental results (section 4.2) show that the learned temporal difference of TDL can effectively boost the segmentation performance quantitatively and qualitatively.

\subsection{Liver Region Guidance} 
\begin{figure}[h]
  \centering
  \includegraphics[width=0.7\columnwidth]{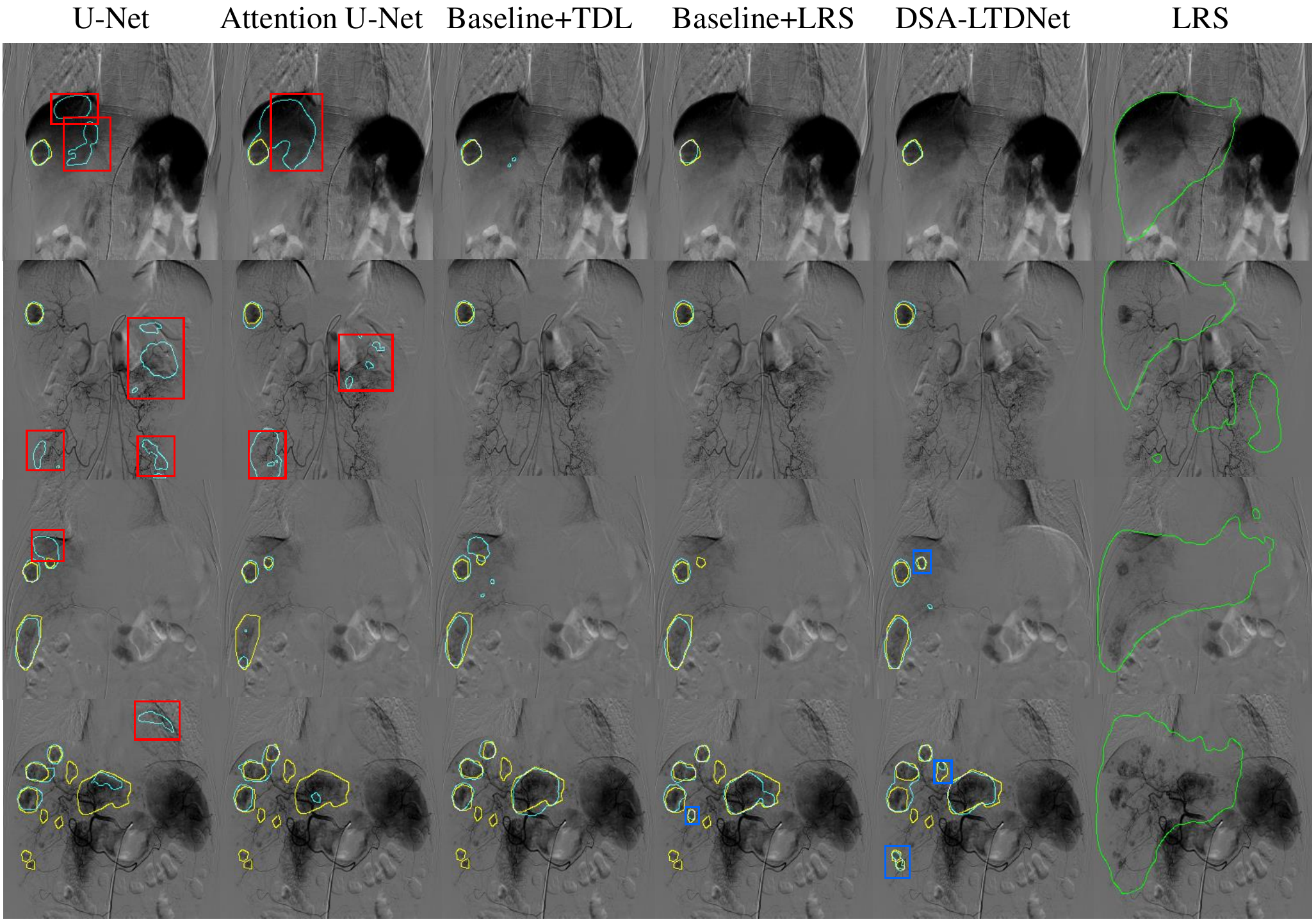}
  \caption{Visualization of the segmentation results by different methods on the testing data. Ground truth and predicted mask of tumors are labeled in yellow and cyan-blue. Predicted masks of liver regions by LRS are labeled in green in the last column. False positive regions and missed tumor regions in predicted mask are marked in red and blue.}
  \label{fig3}
\end{figure}

Our experimental results show that the basic models on the key frame sometimes segment the tumor mask outside the liver region (Fig.\ref{fig3}), while such mistakes do not occur in clinical practice due to their medical knowledge that tumors are always located in or near the liver regions which generally appear at the top left of DSA videos (Fig.\ref{fig1}). Such mistakes happen because some anatomical structures (especially gastrointestinal tissues and healthy liver tissues filled with contrast agent) have highly similar shape, contrast and intensity to tumors in DSA videos. Consequently, the LRS sub-network based on U-Net, is co-trained on the key frames to predict the liver regions. The predicted liver region masks (LM) produced by LRS are plugged the spatial prior knowledge of tumor regions into our network, which successfully guides our model to locate tumors accurately.

\subsection{Architecture and Loss Function}

We integrate the aforementioned module and sub-network into the whole architecture (Fig.\ref{fig2}). Key frames (256 × 256 × 1), along with temporal difference (256 x 256 x 1) learned by LTD and the predicted liver region masks (256 × 256 × 1) by LRS, are concatenated and fed into the final fusion segmentation(FFS) network. We co-train the 3 networks simultaneously. The followings are the loss functions: 

\begin{equation}
L_{LTD}=|I_{LTD}-I_{FD}|_{L1}
\end{equation}
\begin{equation}
L_{LRS} = a * |I_{LRS}-I_{LM}|_{BCE} + (1-a) * |I_{LRS}-I_{LM}|_{DICE}
\end{equation}
\begin{equation}
L_{seg} = a * |I_{seg}-I_{GT}|_{BCE} + (1-a) * |I_{seg}-I_{GT}|_{DICE}
\end{equation}
\begin{equation}
L_{Total}=\lambda_{0} * L_{LTD} + \lambda_{1} * L_{LRS} + L_{seg}
\end{equation}
Where $L_{LTD}$ denotes the loss of LTD; $L_{LRS}$ denotes the loss of LRS; $L_{seg}$ denotes the segmentation loss of tumors;  $L_{Total}$ is the total loss of the whole network );
$a$, $\lambda_{0}$ and $\lambda_{1}$ are 0.5, 0.1 and 1 respectively. 

\section{Dataset Construction}

\subsubsection{Collection}
The DSA data is acquired from 486 HCC patients in the TACE procedure (3 DSA images are acquired form one patient, giving 488 samples in total) with 362 for training and 124 for testing at Hospital A. All of patients have been diagnosed with HCC in BCLC stage B \cite{forner2010current} among January 2010 and December 2019, which was confirmed by biopsy or radiological imaging studies according to the guidelines \cite{bruix2005management}. The included criteria are as followings: 1) over 18 years of age; 2) without any prior treatment or surgery for HCC and any known active malignancy in the past 3 years; 3) Child-Pugh class A or B \cite{durand2005assessment}; 4) Eastern Cooperative Oncology Group (ECOG) \cite{oken1982toxicity} performance scores are 2 or below. The following situations are excluded in the selection: 1) concurrent ischemic heart disease or heart failure; 2) history of acute tumor rupture with hemoperitoneum; 3) Child-Pugh class C cirrhosis; 4) history of hepatic encephalopathy, intractable ascites, biliary obstruction, and variceal bleeding; 5) extrahepatic metastasis, and tumor invasion of the portal vein or branch.

\subsubsection{Annotation}
Each DSA video contains 20-30 frames with 1021×788 pixel-wise resolution. We invited experienced radiologists to annotate the ground truth of the HCC and the corresponding liver regions. We augment the data by selecting 2 more consecutive frames after the key frame for the training, which is similar to key frame because the imaging of tumors keeps stable shortly after the key frame. All of annotations are performed with Labelme platform \cite{russell2008labelme}. There are 760 annotated tumors in total with 212 tumors of poor blood, and 548 tumors of rich blood in our DSA dataset of 488 samples. Note that each patient has up to 10 tumors. The range of tumor size is [6.5, 217] and the average is 69.37 in pixels.

\section{Experiments and Results}
\subsection{Implementation Details}
All the models are trained from scratch based on PyTorch, using a NVIDIA GTX 8119MiB GPU. We do not apply the popular data augmentation (e.g., rotation) due to the special anatomical knowledge of HCC and liver area. In the training settings, the batch size is set to 8 and our proposed network consisted of 3 models is co-trained for 150 epochs. We used the Adam optimizer and the CosineAnnealingLR scheduler to update the network parameters. The initial learning rate is set to 0.001.
In experimenting U-Net, Attention U-Net \cite{oktay2018attention} U-Net++ \cite{zhou2018unet++} and nnU-Net \cite{isensee2018nnu} are trained separately to segment tumors on the key frame at first. At the second stage, we concat the key frame, with one of the followings respectively, which are optical flow, frame difference, background subtraction, learned temporal difference and predicted liver region masks and trained the segmentation networks for comparison. Eventually, we combine the TDL module and the LRS sub-network with the FFS sub-network, and co-trained the DSA-LTDNet.

\begin{table}[ht]

\centering
\caption{Dice scores (\%) for basic models and proposed methods}
\small
\begin{tabular}{ll}
Method & Dice \\
\hline

U-Net (baseline) & 70.75 \\
Attention U-Net & 71.03\\
U-Net++ & 70.73 \\
nnU-Net & 71.89 \\
\end{tabular}
\qquad
\begin{tabular}{ll}
\centering
\small

Method & Dice \\
\hline
FFS + LRS & 72.01 (+ 1.26)\\
FFS + TDL & 72.32 (+ 1.57)\\
DSA-LTDNet (Our Method)&\textbf{73.68 (+ 2.97)}\\

\end{tabular}
\label{tab1}
\end{table}

\begin{table}[ht]
    \centering
    \caption{Dice scores (\%) for TDL comparison experiments}
    \small
    \begin{tabular}{lll}
         Method & Inputs (concat) &  Dice \\
         \hline
         FFS & KF + OF & 68.35 (- 2.40)\\
         FFS & KF + FD & 71.73 (+ 0.98)\\
         FFS & KF + BS & 68.89 (- 1.86) \\
         FFS & KF + TDL w/o supervision & 69.90 (- 0.85)\\
         FFS & KF + TDL w/   supervision & \textbf{72.32 (+ 1.57)}
    \end{tabular}
    \label{tab2}
\end{table}

\subsection{Quantitative Analysis}

The results of basic models and our proposed networks are in Table \ref{tab1}. The left part of Table 1 displays that 4 basic models, U-Net, Attention U-Net U-Net++ and nnU-Net have similar segmentation capabilities on the key frames. That tumors of poor blood supply have low contrast and blurry boundaries in DSA videos, along with other mentioned problems, brings great difficulty to HCC segmentation of basic models. Considering models' complexity of implementation and modification, we choose the easiest U-Net as the backbone of DSA-LTDNet. U-Net's dice score is 70.75\%, which is regarded as our baseline. In the right part, it is worth noting that our proposed auxiliary models substantially improve the performance of HCC segmentation in DSA videos. With LRS and TDL added separately, the results are improved by 1.26\% and 1.57\%. Note that, the proposed method with both TDL and LRS added, achieves the best result, which is higher than U-Net by 2.97\%. These numbers indicate that our methods take remarkable effects. The combination of LRS and TDL is successful as well.\\
At the second stage of our experiments, we tried to use traditional object detection methods as auxiliary information to support the network learn the motion information from DSA videos and segment tumors. The results of experiments in this stage are listed in Table \ref{tab2}. The traditional methods did not work well. The results of optical flow and background subtraction are even reduced by 2.40\% and 1.9\%. It is because that the motion information of other anatomical tissues displays similarly in optical flow and background subtraction as well, which increases the difficulty of final segmentation instead. Afterwards, we used a CNN network to automatically learn the latent temporal motion without any supervision but it did not succeed as expected, whose result is 69.90\%. According to our analysis, the failure is caused by the directionless learning without supervision, which might confuse the TDL network and feedback some noise information to the FFS network. We ceatively propose to use frame difference, which increases the result by 0.98\%, to guide the TDL network. Eventually, this architecture improves the results by 1.57\% compared to the baseline, which satisfies our expectations.

\subsection{Qualitative Results}

Fig. 3 illustrates a visual comparison of the ground truth, 2 basic models, baseline with TDL, baseline with LRS and our DSA-LTDNet on the key frame. It displays that basic models successfully produce accurate segmentation on tumors, but tumors' locations and shapes are not as accurate as our proposed methods. Such observations are particularly apparent in the first 3 rows of Fig. 3. Moreover, our proposed methods can obtain more accurate results by removing false positive regions under the guidance of liver regions and reducing the negative effect of motion artifacts, compared with basic models in the first 3 rows. It is noteworthy that DSA-LTDNet's performance on multiple small tumors marked blue in the last row, is signiﬁcantly better than that of basic models due to the effective learning of temporal difference.

\section{Discussions and Conclusions}
We raise the problem of the HCC segmentation in DSA videos, which is proven significant and challenging in clinical practice. The novel DSA-LTDNet with a dominant TDL module, a supplementary LRS sub-network and a FFS sub-netowrk is proposed. The TDL module can efficiently capture the distinct motion information of changing tumors for delineating tumor regions, while the prior anatomical knowledge of liver regions produced by LRS can support locating tumor regions. We also collect and annotate a relatively large DSA dataset, where we tested our networks. Experimental results show that our networks can effectively learn the latent motion information and achieve the state-of-the-art performance (DICE coefficient: 0.7368) on commonly used evaluation metrics. In the future, in order to improve the robustness, we plan to apply our networks to other DSA datasets from multiple hospitals, which are being collected. We will also facilitate the translation of CNN to clinical practice, especially in the treatment of HCC in the future.

\bibliographystyle{ieeetr} 
\bibliography{ref}

\end{document}